# Comparing Results of Thermographic Images Based Diagnosis for Breast Diseases

É. O. Rodrigues[1], A. Conci[1], T. B. Borchartt[2], A. C. Paiva[2], A. Correa Silva[2], and T. MacHenry[3]
[1] Instituto de Computação, Universidade Federal Fluminense, São Domingos, 24210-240 Niterói-RJ, Brazil
[2] Instituto de Computação, Universidade Federal do Maranhão, Avenida dos Portugueses, 65085-580 São Luiz-MA, Brazil
[3] York University, 4700 Keele Street, ON M3J 1P3, Toronto, Canada
*erickr@id.uff.br, aconci@ic.uff.br*

*Abstract* - This paper examines the potential contribution of infrared (IR) imaging in breast diseases detection. It compares obtained results using some algorithms for detection of malignant breast conditions such as Support Vector Machine (SVM) regarding the consistency of different approaches when applied to public data. Moreover, in order to avail the actual IR imaging's capability as a complement on clinical trials and to promote researches using high-resolution IR imaging we deemed the use of a public database revised by confidently trained breast physicians as essential. Only the static acquisition protocol is regarded in our work. We used 102 IR single breast images from the Pro Engenharia (PROENG) public database (54 normal and 48 with some finding). These images were collected from Universidade Federal de Pernambuco (UFPE) University's Hospital. We employed the same features proposed by the authors of the work that presented the best results and achieved an accuracy of 61.7 % and Youden index of 0.24 using the Sequential Minimal Optimization (SMO) classifier.

*Keywords* – **Breast Cancer; Pattern Recognition; Infrared Imaging, Thermography, Computer-Aided Diagnosis (CAD).**

## I. INTRODUCTION

Clinical decision support systems (CDSSs) assist physicians through the analysis of patient clinical variables; they improve health care quality, reduce health care costs and thus are effective systems regarding medical errors prevention [1]. Computer-Aided Diagnosis (CAD) systems classify organs as with or without any abnormality using medical images as input data [2]. Both computer systems involve some kind of data mining and Pattern Recognition (PR) methodology [3]. In this methodology, first the breast region is separated from the rest of the image (i.e. segmented), then features are extracted and some kind of artificial intelligence (such as machine learning and classification) algorithms are used to classify the organs in analysis as normal or presenting any type of disease [4].

Breast cancer is an important cause of death among women all over the world which, therefore, justifies researches on early diagnosis [5]. Furthermore, infrared thermography has been proven to be a promising technique on early diagnosis of breast pathologies [6- 7]. However, its uses are yet in a research level, more works on the field need to be done in order to establish its real potentialities [8-9]. This work aims to contribute to this area by comparing results and approaches with regard to Pattern Recognition for Computer-Aided Diagnosis (CAD) and Clinical Decision Support Systems (CDSSs) [10-11]. Moreover, we enhance previously published works by applying a known method to a public database and comparing the obtained results [12, 13]. This paper is organized in five sections. Section 2 does an overview of the most recent works involving Infrared Imaging for breast cancer diagnosis. Section 3 presents some aspects of the applied approach. Section 4 considers a comparison of works. The last section addresses the reasons of doing this work and our conclusion.

## II. RELATED WORKS

Breast infrared imaging aggregates two main issues: low sensitivity for deep and small masses and problems on interpretation of the meaning of "hot spots" (it can be a local inflammation, an important finding for diagnostic purposes, or even a normal vein) [14]. To overcome such disadvantages, the use of pattern recognition techniques is fundamental.

Subsequently to the segmentation of the breast region, the next step on using pattern recognition techniques is related to choosing relevant characteristics that should be extracted from the image [3]. These characteristics, i.e., features, compose a vector of values used for the classification of an arbitrary breast as a healthy or an unhealthy one. Since a very recent work addresses a review on works related to feature extraction as well as a comparison of some papers regarding pattern recognition for breast cancer diagnosis; in this section, we address only works published subsequently in relation to this recent one [4]. Figure 1 depicts the general steps and the basic workflow applied on this work.

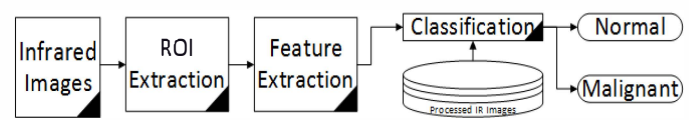

Figure 1. Basic workflow of the steps applied on this work.

One of these new works considers 50 infrared (IR) breast images (25 normal and 25 cancerous) collected from Singapore General Hospital, Singapore [10]. The chosen features are based on texture, using co-occurrence and run-length matrix. Conclusively, four features: first moment, third moment, run percentage and gray level non-uniformity were selected to feed the classification step where the Support Vector Machine (SVM) classifier was applied.

The features used in [15] are based on statistical measures, histogram, Higuchi´s fractal dimension and three geo statistical methods: Geary coefficient, Moran's index and Ripley's K function. So that 14 IR images of healthy patients and 37 images of patients with some pathology were used. Regarding the patients with any pathology there are 74 single breasts (1 lipoma, 1 abscess, 2 steatonecrosis, 7 cysts, 8 fibroadenomas, 10 carcinomas, 19 unclassified tumors and 26 healthy breats). The SVM algorithm was used for classification and the genetic algorithm was used to optimize the results. The results achieved by these two previous listed works are better described in Section 4.

We contacted the authors of [10] considering the use of their images. However, they mentioned that for ethical reasons, its access is private. In order to have a fair comparison we extracted the same features proposed by Acharya et al. [10] from the images employed on the work of Borchartt et al. [15] that are available at the PROENG public database [16].

### III. Aspects of the Applied Approach

We have used 102 IR single breast images from the PROENG public database (54 normal and 48 with some finding). These images are collected from UFPE University's Hospital and published under approval of the ethical committee whereas every patient must sign consent [17].

The segmentation of the breast region, i.e., the extraction of the region of interest (ROI), regarding thermographic images, is an especially very hard task due to their amorphous nature and lack of clear limits [18]. In our approach we have used the same methodology that composed the automatic ROI segmentation developed in [19] that was proven to produce good results regarding a comparison to a ground truth created by a group of specialists [17, 20]. The lines in Figure 2 show the border of some already extracted ROIs. Posteriorly to the ROI extraction, we have computed the four features applied in [10].

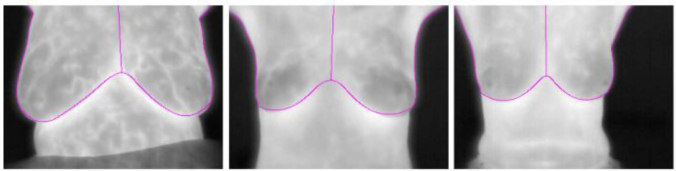

Figure 2. Samples of ROI limits of the images employed on this work.

For every breast image, the features: first moment, third moment, gray level non uniformity and run percentage were extracted. In emails from the authors of [10], we learn that only the distance parameter $\Delta x, \Delta y = \{0,1\}$ is considered. The chosen distance parameter is not clear on the original paper. Hence, by knowing it, we are able to apply the same informed distance parameter in our approach.

Furthermore, to compute the co-occurrence matrix, we iterate through every pixel $p_1$ of the image and ask if a pixel $p_2$ at a distance $(\Delta x, \Delta y)$ away from $p_1$ is equal to the gray level of the pixel $p_1$, if so, we increment a score associated to these two pixels ($p_1$ and $p_2$) in relation to the distance parameter. Hence, we compute the matrix using the Equation (1) where, given an arbitrary image $I$ that is $m$ x $n$ pixels wide, $C_{\Delta x, \Delta y}(i,j)$ denotes the co-occurrence score of a gray value $i$ associated to a gray value $j$. The function $V(x,y)$ denotes the gray value of a pixel on the position $(x,y)$ within an arbitrary image $I$ [10].

$$C_{\Delta x, \Delta y}(i,j) = \sum_{x=1}^{n} \sum_{y=1}^{m} \begin{cases} 1, if\ V(x,y) = i\ \text{and}\ V(x+\Delta x, y+\Delta y) = j \\ 0, otherwise \end{cases} \quad (1)$$

The Equation (2) consists of returning the probability of a gray level $i$ to co-occur with a gray level $j$ among every co-occurrence for any pair of gray levels on the whole co-occurrence matrix. The Equation (3) is used to compute the moment, where $g$ is the moment degree. We only consider the first and the third ones as well as Acharya et al. do (i.e. $g = 1$ and $g = 3$) [10].

$$P_{(\Delta x, \Delta y)}(i,j) = \frac{C_{\Delta x, \Delta y}(i,j)}{\sum C_{\Delta x, \Delta y}(i,j)} \quad (2)$$

$$m_g = \sum_i \sum_j P_{\Delta x, \Delta y}(i,j)\,(i-j)^g \quad (3)$$

To compute the gray level non uniformity feature we must introduce the function $R_\theta(t,l)$ which represents the run length value of an arbitrary image $I$ given a $\theta$ parameter. The run length function consists of returning the number of times a gray level $i$ continuously appears in the image with a length $l$, respecting a $\theta$ direction. Thus, to compute the feature we iterate once through every gray level of the image and for each one we iterate through every possible length (regarding the worst case scenario, the maximum length is equal to the diagonal of the image). Accordingly, on equation (4), we depict the function $G_\theta$ responsible for computing the gray level non uniformity feature given a $\theta$ parameter [10].

$$G_\theta = \frac{\sum_i (\sum_j R_\theta(i,j))^2}{\sum_i \sum_j R_\theta(i,j)} \quad (4)$$

The third feature to be extracted is the run percentage, which consists of summing all the elements of the run length matrix and dividing it by the area $A$ of the image as shown on equation (5) by the function $RP_\theta$ [10].

$$RP_\theta = \frac{\sum_i \sum_j R_\theta(i,j)}{A} \quad (5)$$

Acharya et al. [10] also did not mention which one or if all directions were chosen for the $\theta$ parameter, or even how they were combined. We then decided to extract all the four possible directions (0°, 45°, 90°, 135°) for both features (gray level non uniformity and run percentage), and hence, educing 8 values in total. Moreover, we also include results from $m_1$ and $m_3$, so, the final output has 10 elements in the feature vectors. Moreover, we chose some classifiers that were prone to give the best results.

Table I shows some statistics for some of these chosen classifiers. The k-fold stratified cross validation technique was applied to every classifier. When the k variable of the k-fold technique is equal to the number of instances, it may be called the *leave one out* technique. We have empirically chosen some k-range for all algorithms and focused on the classifiers that returned the higher accuracy.

TABLE I. ACCURACY OF CLASSIFIERS APPLIED ON THIS WORK

| Returned Classifier Accuracy | | | | | |
|---|---|---|---|---|---|
| Classifier | Accuracy | Sens. | Spec. | Youden | Test Mode |
| SMO[a] | 61.8% | 62.9% | 61.8% | 0.24 | 7-fold cross validation |
| RBFNetwork[b] | 58.5% | 60% | 58.8% | 0.18 | leave one out |
| NaiveBayes[c] | 56.8% | 56.8% | 56.9% | 0.12 | 7-fold cross validation |
| SVM[d] | 53.9% | 50% | 57.4% | 0.07 | leave one out |

a. C: 1.0, Kernel: Polykernel, regOptimizer: regSMO
b. minStdDev: 0.27, numClusters: 1
c. Standard Parameters
d. Svm: NU, Kernel: RBF, Gamma: 0.00015, NU: 0.09

The images used are listed in the appendix B of [15]. The open code software LibSVM [21] and Weka Data mining software [22] were used in our experiments.

IV. COMPARISON WITH OTHER RESULTS

The distinction between healthy and unhealthy breast image is a binary classification problem. Four possible cases are considered: *true negative* (TN), where a healthy breast is correctly classified as healthy; *true positive* (TP), which corresponds to a diseased breast that is correctly classified as unhealthy; *false positive* (FP), where a healthy breast is incorrectly classified as unhealthy; and *false negative* (FN), which corresponds to an unhealthy breast that is incorrectly identified as healthy. Using these four possible results, i.e., the number of instances in each one of these four classes, different methodologies may have their performance compared by the use of measures depicted by the following Equations (6-9):

$$accuracy = \frac{TP+TN}{TP+FP+FN+TN} \quad (6)$$

$$precision = \frac{TP}{TP+FP} \quad (7)$$

$$sensitivity = \frac{TP}{TP+FN} \quad (8)$$

$$specificity = \frac{TN}{TN+FP} \quad (9)$$

Combinations of these measures are also a simpler way to compare the performance of different systems. Particularly, one of the most used representations is the Youden's index shown on Equation (10).

$$sensitivity + specificity - 1 \quad (10)$$

Some works on the analysis of breast thermal images provide classification results using the accuracy, specificity and sensitivity measures and sometimes present the corresponding area under curve (AUC) of their methods as measures of robustness [23]. Results of [10, 24-26] consider normal breast and malignant breast cancer. The works [6, 27, 28] consider benign and malignant breast cancer [15]. Our work considers normal and various degrees of abnormal conditions on the studied breast along with an approach that originally was applied to either malignant or healthy breasts. Table II summarizes the results of all the most recent works cited. The numbers within the brackets mean the ratio between healthy breasts per unhealthy ones of each work.

TABLE II. COMPARISON OF RESULTS

| Work | Samples used | Sens. | Spec. | Acc. | Y index |
|---|---|---|---|---|---|
| Arora *et al.* (2008) | 94 (34/60) | 97% | 44% | - | 0.41 |
| Wishart *et al.* (2010) | 106 (41/65) | 48% | 70% | - | 0.18 |
| Umadevi *et al.* (2010) | 50 (44/6) | 66.7% | 97.7% | - | 0.64 |
| Acharya *et al.* (2012) | 50 (25/25) | 85.7% | 90.5% | 88.1% | 0.76 |
| Brochatt (2012)[a] | 51 (14/37) | 83,8% | 57,1% | 76,5% | 0,41 |
| Brochatt (2012)[b] | 51 (14/37) | 83.8% | 78.6% | 82.4% | 0.62 |
| Brochatt (2012)[c] | 51 (14/37) | 91.9% | 78.6% | 88.2% | 0.71 |
| This work (SMO Result) | 102 (54/48) | 61.72% | 62.9% | 61.8% | 0.24 |

a. Original work.
b. Moran Index
c. Optimized version

V. CONCLUSION

A major factor related to the inconsistency of the comparison among works is the use of distinct IR images. We consider here a public database where the region of interest (ROI) of each image was compared to a ground truth created by a group of experts [16, 17]. We do also employ images with various degrees of abnormal conditions instead of just malignant or just healthy.

Even though applying the same approach of Acharya et al. [10], we were not able to reach their results. In our case, the best result came from Sequential Minimum Optimization (SMO) classifier, which is based on the SVM algorithm and on an intrinsic optimization. In conclusion, our approach clearly evince all the details of each extracted feature, employs an available public database during the extraction of these features and uses open source software code [22] which allows other authors to reproduce and compare new results with ours.

Moreover, we consider the diagnosis by breast (not per patient), as being more realistic.

ACKNOWLEDGMENT

This research was supported by the Brazilian CAPES projects: Casadinho/ProCad, and ProCad-NF 540/2009. A. Conci wants to thank the CNPq project 302298/2012-6. E. O. Rodrigues wants to thank CAPES for the concession of master's stipend. T. Borchartt wants to thank CNPq for the concession of post doc's scholarship by means of INCT-MACC.